\documentclass[amsmath,amssymb,prl,aps,showpieces,superscriptaddress,twocolumn]{revtex4-1}
\usepackage{amsmath,amsfonts,amssymb,amsthm,graphics,graphicx,epsfig,bbm}
\usepackage[colorlinks=true,citecolor=blue,linkcolor=blue,urlcolor=blue]{hyperref}
\usepackage[usenames]{color}
\usepackage{graphicx}
\usepackage{subfigure}
\usepackage{amsmath}
\usepackage{epsfig}
\usepackage{dcolumn}
\usepackage{bm}
\usepackage{color}
\usepackage{times}
\usepackage{epstopdf}
\usepackage{amssymb}
\usepackage{amstext}
\usepackage{latexsym}
\usepackage{hyperref}
\usepackage{amsfonts}
\usepackage{psfrag}
\usepackage{soul,xcolor}
\usepackage[normalem]{ulem}
\usepackage{dsfont}
\usepackage{txfonts}
\usepackage{float}
\usepackage{algorithm}
\usepackage{algpseudocode}

\newcommand{\ket}[1]{\vert #1 \rangle}

\usepackage{tikz,xcolor,hyperref}
\definecolor{lime}{HTML}{A6CE39}
\DeclareRobustCommand{\orcidicon}{%
	\begin{tikzpicture}
	\draw[lime, fill=lime] (0,0) 
	circle [radius=0.16] 
	node[white] {{\fontfamily{qag}\selectfont \tiny ID}};
	\draw[white, fill=white] (-0.0625,0.095) 
	circle [radius=0.007];
	\end{tikzpicture}
	\hspace{-2mm}
}

\foreach \x in {A, ..., Z}{%
	\expandafter\xdef\csname orcid\x\endcsname{\noexpand\href{https://orcid.org/\csname orcidauthor\x\endcsname}{\noexpand\orcidicon}}
}

\begin{document}

\setstcolor{red}

\title{Noise reduction via optimal control in a light-matter quantum system} 
\date{\today}

\author{Francisco Albarr\'an-Arriagada \orcidA{}} 
\affiliation{Departamento de F\'isica, CEDENNA, Universidad de Santiago de Chile (USACH), Avenida V\'ictor Jara 3493, 9170124, Santiago, Chile.}


\author{Guillermo Romero \orcidB{}}
\email[G. Romero]{\qquad guillermo.romero@usach.cl}
\affiliation{Departamento de F\'isica, CEDENNA, Universidad de Santiago de Chile (USACH), Avenida V\'ictor Jara 3493, 9170124, Santiago, Chile.}

\author{Enrique Solano \orcidD{}}
\affiliation{Kipu Quantum,  Greifswalderstrasse 226, 10405 Berlin, Germany}

\author{Juan Carlos Retamal \orcidC{}}
\affiliation{Departamento de F\'isica, CEDENNA, Universidad de Santiago de Chile (USACH), Avenida V\'ictor Jara 3493, 9170124, Santiago, Chile.}

\begin{abstract}
Quantum noise reduction below the shot noise limit is a signature of light-matter quantum interaction. A limited amount of squeezing can be obtained along the transient evolution of a two-level system resonantly interacting with a harmonic mode. We propose the use of optimal quantum control over the two-level system to enhance the transient noise reduction in the harmonic mode in a system described by the Jaynes-Cummings model. Specifically, we propose the use of a sequence of Gaussian pulses in a given time window. We find that the correct choice of pulse times can reduce the noise in the quadrature field mode well below the shot noise, reaching reductions of over 80$\%$. As the Jaynes-Cummings model describes a pivotal light-matter quantum system, our approach for noise reduction provides an experimentally feasible protocol to produce a non-trivial amount of squeezing with current technology.

\end{abstract}

\maketitle

\emph{Introduction}.\textemdash
Squeezed states \cite{Walls1983Nov,BibEntry,Loudon1987Jun,Leuchs1988May,Teich1989Dec,Dodonov2002Jan,Reid2009Dec,Lvovsky2015Jan,Chekhova2015Feb,Andersen2016Apr} refer to quantum states that exhibit reduced uncertainty in one of the conjugate variables, such as position and momentum or phase and amplitude. In quantum optics, squeezed states have garnered significant attention due to the application to gravitational wave detection~\cite{Caves1981Apr, BibEntry2011Dec, Aasi2013Aug}.  Additionally, squeeze states are an important resources for several tasks in quantum technologies such as quantum teleportation~\cite{Braunstein1998Jan,Furusawa1998Oct}, continuos variable quantum computing~\cite{Menicucci2006Sep}, and quantum error correction~\cite{Aoki2009Aug,Lassen2010Oct}. The process of creating and manipulating these states, known as squeezed states engineering~\cite{Mehmet2018Dec,Magana-Loaiza2019Sep,Barral2020Oct,Grimsmo2017Jun,Vahlbruch2007Oct,Hastrup2021Apr,Groszkowski2022Jan}, involves deliberately altering the quantum fluctuations to enhance precision measurements improving the performance of quantum devices. Researchers are actively exploring various methods for generating and manipulating squeezed states, including nonlinear optical processes, parametric down-conversion, and interactions in atomic systems \cite{Andersen2016Apr}. These methods enable the controlled creation of squeezed states, paving the way for quantum optics and quantum information science advancements.

A quantum field interacting with a two-level system (TLS), described by the Jaynes-Cummings model (JCM) \cite{Jaynes1963Jan,Eberly1980May,Narozhny1981Jan,Shore1993Jul,BibEntry2006Oct,Larson2021Dec}, can experience a squeezing process during transient evolution, but the extent of reduction is limited, regardless of the initial state \cite{Meystre1982Jun,Kuklinski1988Apr}. The amount of transient squeezing can be enhanced through collective effects increasing the number of atoms~\cite{Retamal1997Mar}. An alternative strategy for increasing transient noise reduction relies on applying a periodic sequence of kicks to the two-level system~\cite{Retamal1994Aug}. The search for non-classical states in this system continues to be an exciting problem \cite{Abah2020May,Uria2020Aug}. The best strategy to reach maximal squeezing in a light-matter system is still an open question, where optimal control over the TLS seems promising.

Quantum control plays a pivotal role in quantum information processing, encompassing quantum computation~\cite{Machnes2018Apr,Hegade2021Feb}, quantum communication~\cite{Zwerger2012Jun,Rubino2021Jan}, quantum sensing~\cite{Rembold2020Jun,Poggiali2018Jun}, and quantum simulation~\cite{Iram2021Jan,Sels2017May}. In this context, manipulating TLS through pulses to generate coherent operations such as rotations and superpositions is at the heart of quantum technologies. The generation of Gaussian shape pulses is one of the most fundamental types of control and, therefore, feasible in most current quantum platforms \cite{Ciani2022Nov,Krantz2019Jun}. Researchers continuously explore new control techniques, improving the accuracy and reliability of quantum operations, heralding a new era of quantum-enhanced applications with light-matter interaction. Optimizing quantum control protocols is a powerful tool for finding optimal sequences and shapes for the control fields to reduce a cost function among a large set of input parameters.

In this work, we search for optimal Gaussian pulse sequences to control the TLS that enhance the quantum mode transient squeezing in the JCM. Specifically, we consider the system dynamics by applying a Gaussian pulse train to control $x$-axis rotation in the TLS. These Gaussian pulses change the global system state in the Hilbert space depending on the time the pulses are applied. Our results show a noise reduction larger than 80$\%$ for different pulse widths and initial states by optimizing the pulse time only. This work provides a path to get non-trivial squeezing using only Gaussian control over a TLS, which is common in almost all quantum platforms, opening the door to larger degrees of squeezing by using a general control over the TLS in a JCM.

\emph{Light-matter interaction}.\textemdash
The interaction of a two-level system with one harmonic mode is described by the Quantum Rabi model \cite{Braak2011Aug,Braak2016Jun} given by:
\begin{equation}
H=\hbar \omega a^{\dagger}a+\hbar \frac{\omega_0}{2}\sigma_z +\hbar g\sigma_x(a+a^{\dagger})
\end{equation}
where $\omega$ is the mode frequency, $\omega_0$ is the two level frequency, $a$ and $a^{\dagger}$ describes the mode creation and annihilation operators, and $\sigma_k$ represent the $k$-pauli matrix that describes the two level system. This model can be simplified when $g/\omega \ll 1$, and $\omega_0\sim\omega$ to the JCM, which reads
\begin{equation}
H=\hbar \omega a^{\dagger}a+\hbar \frac{\omega_0}{2}\sigma_z +\hbar g(\sigma_- a^{\dagger}+\sigma_+ a),
\end{equation}
where $\sigma_+$ and $\sigma_-$ describes the two level raising and lowering
operators.

Among the features of this bipartite interaction, one of the most impressive is the transient squeezing in field quadratures, as, for example, $X$ quadrature \cite{Meystre1982Jun}. To understand this, we should recall the evolution described by this model in the case of an initial coherent state. Assuming the two-level system initially in the excited state, we have that after a time $t$; the state can be written as $\mid \Psi (t)\rangle = \mid e\rangle \mid \Psi_e \rangle +\mid g\rangle \mid \Psi_g \rangle$ where:
\begin{eqnarray}
\mid \Psi_e \rangle &=& \sum_{n=0} c_{n}\cos {\big(g t\sqrt{n+1}\big)}\mid n\rangle \nonumber \\
\mid \Psi_g \rangle &=& -i\sum_{n=1} c_{n-1}\sin {\big(g t\sqrt{n}\big)}\mid n\rangle 
\label{states1}
\end{eqnarray}
where $c_{n}=e^{-|\alpha|^2/2}\alpha^{n}/\sqrt{n!}$ stands for the coherent state amplitudes. As long as the coherent state amplitude is enlarged, the probability distribution for the coherent state can be well described by a Gaussian distribution 
$e^{-\frac{1}{2\bar{n}}(n-\bar{n})^2}/\sqrt{2\pi \bar{n}}$, where $\bar{n}=|\alpha|^2$~\cite{Gea-Banacloche1991Nov,Gea-Banacloche1993Mar}. In large $\bar{n}$ approximation, the states in Eq.~(\ref{states1}) can be written as $\mid \Psi_e \rangle = \frac{1}{2}(|\psi_{+}\rangle+|\psi_{-}\rangle)$ and $\mid \Psi_g \rangle = \frac{1}{2}(|\phi_{+}\rangle+|\phi_{-}\rangle)$ where
\begin{equation}
|\psi_{\pm}\rangle= \sum_{n=0} c_{n}e^{\pm ig t\left[\sqrt{\bar{n}+1}+\frac{n-\bar{n}}{2({\bar{n}+1})^{1/2}}-\frac{(n-\bar{n})^2}{8({\bar{n}+1})^{3/2}}+\frac{(n-\bar{n})^3}{16({\bar{n}+1})^{5/2}}\right]} |n\rangle 
\label{states3}
\end{equation}
\begin{equation}
|\phi_{\pm}\rangle= \mp i\sum_{n=0} c_{n-1}e^{\pm ig t\left[\sqrt{\bar{n}}+\frac{n-\bar{n}}{2({\bar{n}})^{1/2}}-\frac{(n-\bar{n})^2}{8({\bar{n}})^{3/2}}+\frac{(n-\bar{n})^3}{16({\bar{n}})^{5/2}}\right]}  |n\rangle
\label{states3}
\end{equation}
where we have considered the expansion in Eq.~(\ref{states1}) up to third order in $(n-\bar{n})/\sqrt{\bar{n}+1}$.

This suggests that squeezing in a light-matter quantum system is a competition between nonlinearity and quantum interference. Indeed, squeezing appears in the $X$ quadrature since an interference effect, which is produced because states $(|e\rangle|\psi_\pm\rangle \mp i|g\rangle|\phi_\pm\rangle)/2$ rotate counterclockwise and clockwise in the phase space, generating interference in field state amplitudes~\cite{Retamal1997Mar}. After this interference disappears, the overlap among states vanishes, and fluctuations increase.

\emph{Controlled light-matter systems}.\textemdash Controlling the statistical properties for the field mode can be accomplished by manipulating the two-level system. We are mainly concerned with controlling quantum fluctuations within a time scale shorter than decoherence scales. As we will discuss in short, this is particularly interesting for superconducting circuit platforms, where the stability of qubit and harmonic mode goes to hundreds of microseconds ($\mu s$), which is much longer than a typical time for Rabi oscillations ($\sim 10$ ns). These numbers allow us to consider the Hamiltonian evolution for reduced times $gt\le10$ without the considerable effect of the dissipative process. 

As the fluctuation reduction is produced by the interference of the states described in Eq. (\ref{states1}), one strategy to enhance the reduction is performing a bit-flip operation in the TLS each time the transient squeezing is maximal. It is observed that fluctuation reduction increases using this naive strategy. This effect comes from a quantum interference reinforcing. To see this notice that for large $\bar n$ we have that $|\psi_\pm\rangle\approx |\phi_\pm\rangle$, then $(|e\rangle|\psi_\pm\rangle \mp i|g\rangle|\phi_\pm\rangle)/2 \approx (|e\rangle \mp i|g\rangle|)|\psi_{\pm}\rangle/2 $, we understand that  $(|e\rangle - i|g\rangle|)|\psi_{+}\rangle/2 $ rotates counterclockwise, whereas $(|e\rangle + i|g\rangle|)|\psi_{-}\rangle/2 $ rotates clockwise in phase space. A flip on two level states leads with $(|e\rangle - i|g\rangle|)|\psi_{+}\rangle/2 \rightarrow -i(|e\rangle + i|g\rangle|)|\psi_{+}\rangle/2$ that will rotate clockwise and $(|e\rangle + i|g\rangle|)|\psi_{-}\rangle/2 \rightarrow i(|e\rangle - i|g\rangle|)|\psi_{-}\rangle/2 $ will rotate counterclockwise, recovering in this way the overlap among field states. The amount of squeezing using this strategy is about $60\%$ below the shot noise~\cite{Retamal1994Aug}.  The main question arising at this point is whether other control strategies allow the reduction of fluctuations further. We find that the answer is positive, and the strategy that allows us to find an enhanced reduction relies on an optimization process to find the set of times to act on the TLS to reinforce quantum interference. To this purpose, we consider the TLS to be controlled by a train of Gaussian pulses as follows:
\begin{equation}
H=\hbar \omega a^{\dagger}a+\hbar \frac{\omega_0}{2}\sigma_z +\hbar \lambda(\sigma_- a^{\dagger}+\sigma_+ a)+\sum_{n=1}^{N}\hbar \Omega_i (t)\sigma_x
\end{equation}
where $\Omega _i(t)$ is a Gaussian pulse, centered at time $t_i$, width $\sigma$,  amplitude $\Omega_0$ and central frequency $\omega_p$ given by:
\begin{equation}
\Omega_i (t)= \Omega_0 e^{-\frac{(t-t_i)^2}{2\sigma^2}}\cos(\omega_p t),
\label{Gpulse}
\end{equation}
where we can choose conveniently the value of $\Omega_0$. In what follows, we are concerned with searching for a convenient train of pulses applied on the TLS, which could lead us to squeeze in the quantum mode. 

\begin{figure}[t]
	\centering
	\includegraphics[width=0.8
	\linewidth]{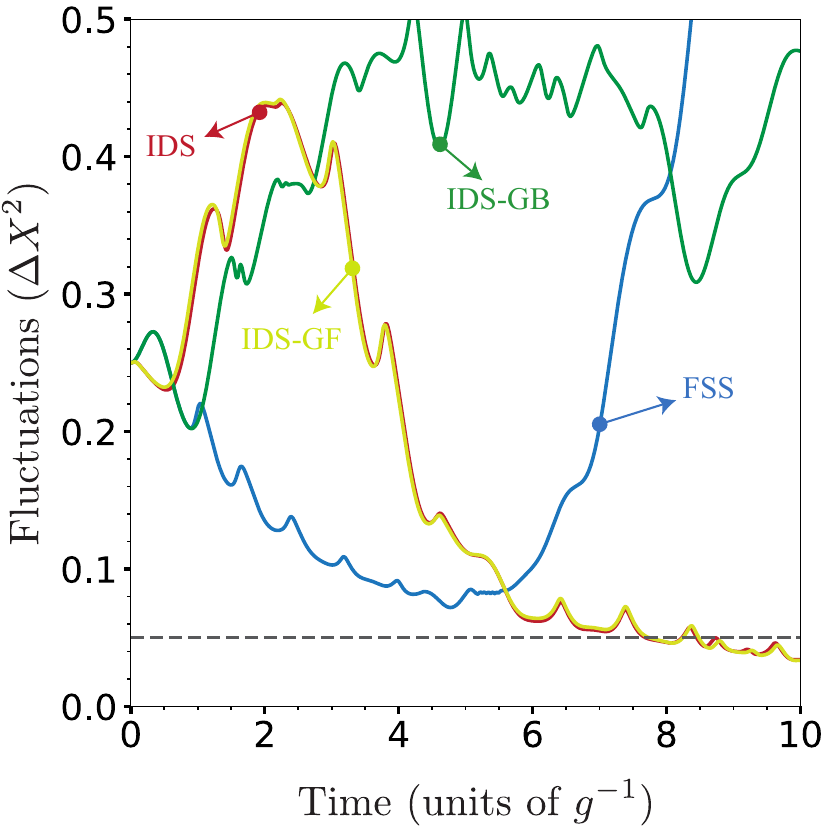}
	\caption{Fluctuations as a function of time given by the JC model and Gaussian pulses obtained with different strategies. The dashed line represents the reduction percentage $\mathcal{R}\%=80\%$, see Eq.(\ref{reductionper}) in the main text.} 
	\label{Fig02}
\end{figure}
\begin{figure}[b]
	\centering
	\includegraphics[width=1\linewidth]{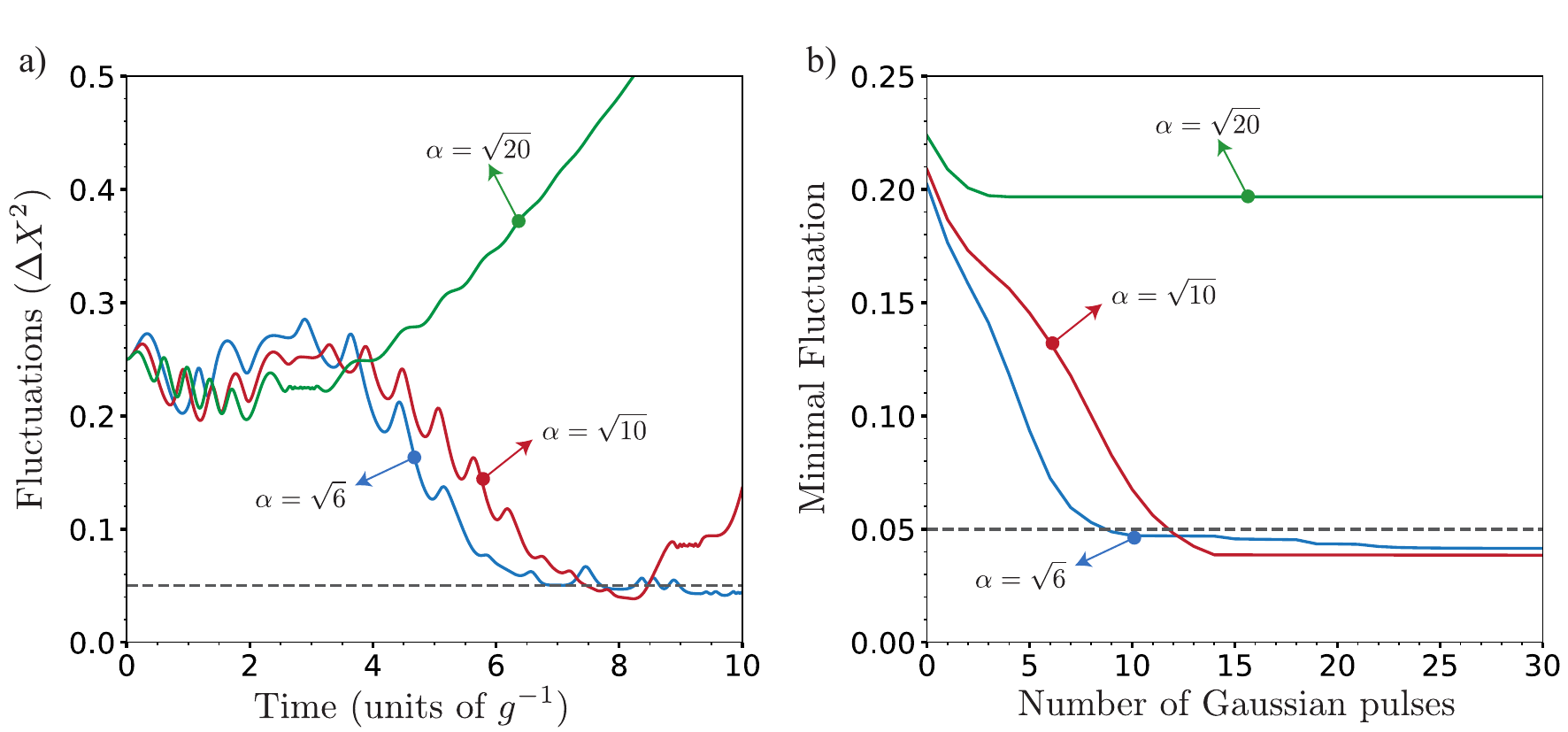}
	\caption{Fluctuation dynamics for coherent states with $\alpha=\sqrt{6}$, $\alpha=\sqrt{10}$, and $\alpha=\sqrt{20}$. We consider for the Gaussian pulses $g\sigma=0.1$. a) Time evolution of cuadrature fluctuation $\Delta X^2$ using 30 Gaussian pulses. b) Minimal value of fluctuation reached with different number of Gaussian pulses.} 
	\label{Fig03}
\end{figure}

\emph{Numerical Results}.\textemdash As described above, the interference between the states described in Eq. (\ref{states1}) plays a central role in reducing the quantum fluctuation. Nevertheless, such interference may or may not be beneficial for reducing quantum fluctuations. For this reason, different minimization strategies will have very different performance. In particular, we will explore, as first instances, four different minimization strategies that are forward sequential strategy (FSS), iterative disorder strategy (IDS), gradient-based iterative disorder strategy  (GB-IDS), and gradient-free iterative disorder strategy (GF-IDS). Each of these strategies provides different trajectories in the Hilbert space. We are concerned with those trajectories exhibiting squeezing in the field $\hat{X}$ quadrature. We consider the Gaussian pulse central frequency to be in resonance with the field mode and the TLS. We choose a field frequency larger than the coupling constant, such as $\omega /g =100$. We consider a time evolution in terms of the adimensional time $gt \in [ 0,10 ]$ and the adimensional pulse width $g\sigma=\{0.025,\, 0.05,\,0.1\}$. 

\emph{Forward sequential strategy.}\textemdash  In the FSS, we search for pulses forward in time. That is, we suppose that we have found the $k-1$ pulses centered in times $t_1<t_2<...<t_{k-1}$, respectively. To find the time $t_k$, where the $k$th pulse is centered, we integrate the  Schr\"{o}edinger equation for $gt$ in the interval $[0,10]$ applying the previous (k-1) pulses plus the k$th$ pulse fixed in $t_k^*\in[t_{k-1},10]$ ($t_0=0$). Once the integration is done, we can calculate the fluctuations in the range $[0,10]$, where its minimal value in that time window is denoted by $f_k^*$.  Each election of $t_k^*$ will give a different value for $f_k^*$. To find the optimal value of $t_k^*$ for which $f_k^*$ is minima, that is, the value of $t_k$; we can use the simplest approach, which is to use a grid time for the desired range of times, evaluate each time-point of the grid and select as the optimal time $t_k$. With this method, we find forward sequential times for the Gaussian pulses. It is important to mention that this strategy is fast since we are doing sequential one-dimensional optimization, even if such optimization is done by brute force.

\emph{Iterative disorder strategy.}\textemdash In the IDS, we suppose again that we know the set of times for the optimal $k-1$ pulses, that is $T_{k-1}=\{t_1,\dots,t_{k-1}\}$, to add a new pulse we consider a new time $t_{k}^*\in [0,10]$, to apply the $k$th pulse obtaining a minimal fluctuation $f^*$. In this part, the difference with the FSS is that we allow that the $k$th is centered before the pulse $k-1$. Next, we select the optimal time for the pulse as in the previous case. As in this method, $t_k$ can be less than $t_{k-1}$ then we perform the next iterative procedure to converge with all the times. Once $t_k$ is found, we replace the first pulse with a pulse centered in $t_1^{*}\in[0,10]$, and in the same way previously described, we find the optimal new time for the pulse one; next, we do the same with the second pulse, and so on until the $k$th one. Next, we evaluate if the new minimal fluctuation is less than $f^*$. If the iteration reduces the fluctuations, we start a new one, which means changing the pulse time of each pulse one by one. We repeat this iterative procedure until the fluctuations between two iterations do not change appreciably. In our case, our tolerance was fixed in $10^{-5}$. In this case, as in FSS, the optimization is done by brute force, defining a time grid.

\begin{figure}[t]
	\centering
	\includegraphics[width=1\linewidth]{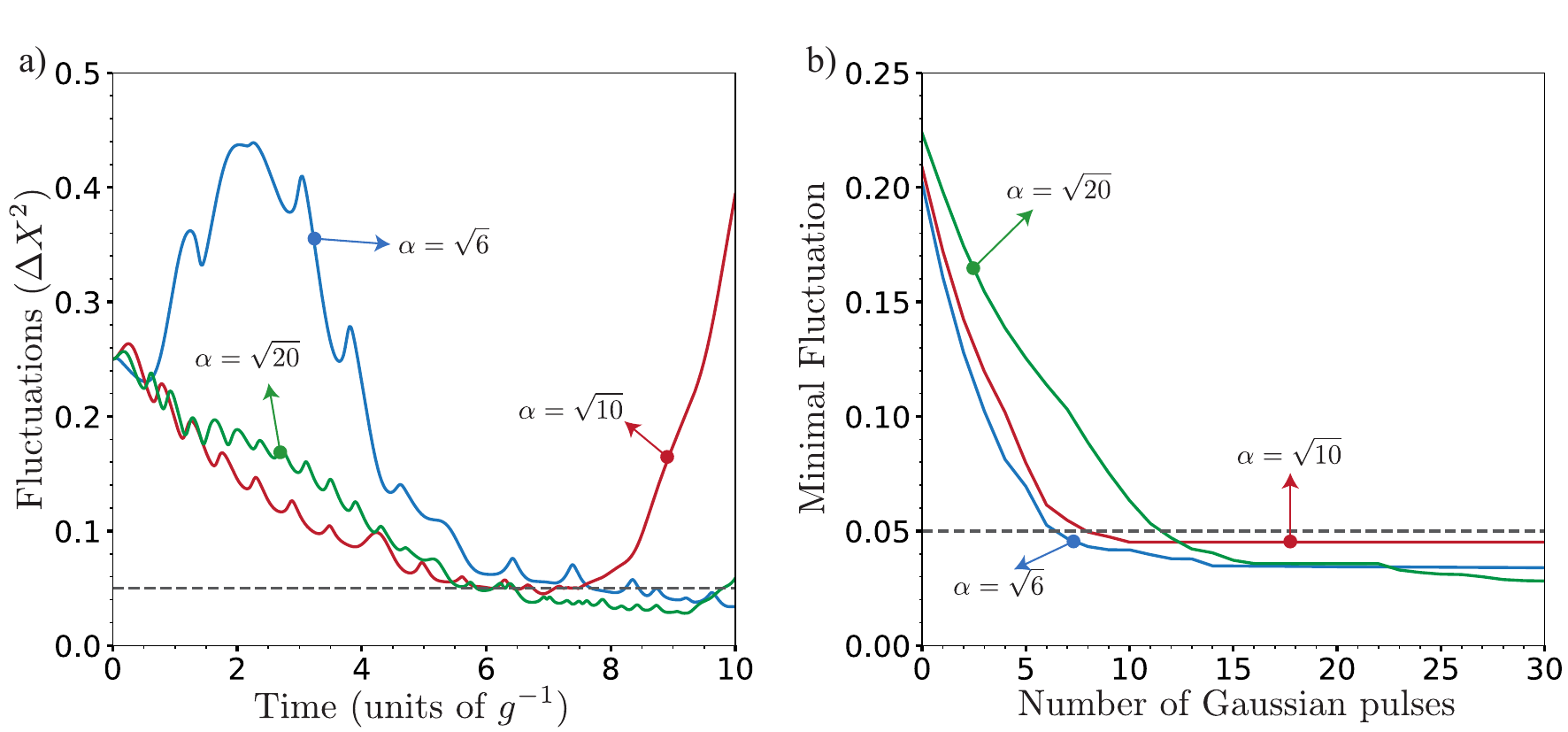}
	\caption{Fluctuation dynamics for coherent states with $\alpha=\sqrt{6}$, $\alpha=\sqrt{10}$, and $\alpha=\sqrt{20}$. We consider for the Gaussian pulses $g\sigma=0.05$. a) Time evolution of cuadrature fluctuation $\Delta X^2$ using 30 Gaussian pulses. b) Minimal value of fluctuation reached with different number of Gaussian pulses.} 
	\label{Fig04}
\end{figure}
\begin{figure}[b]
	\centering
	\includegraphics[width=1\linewidth]{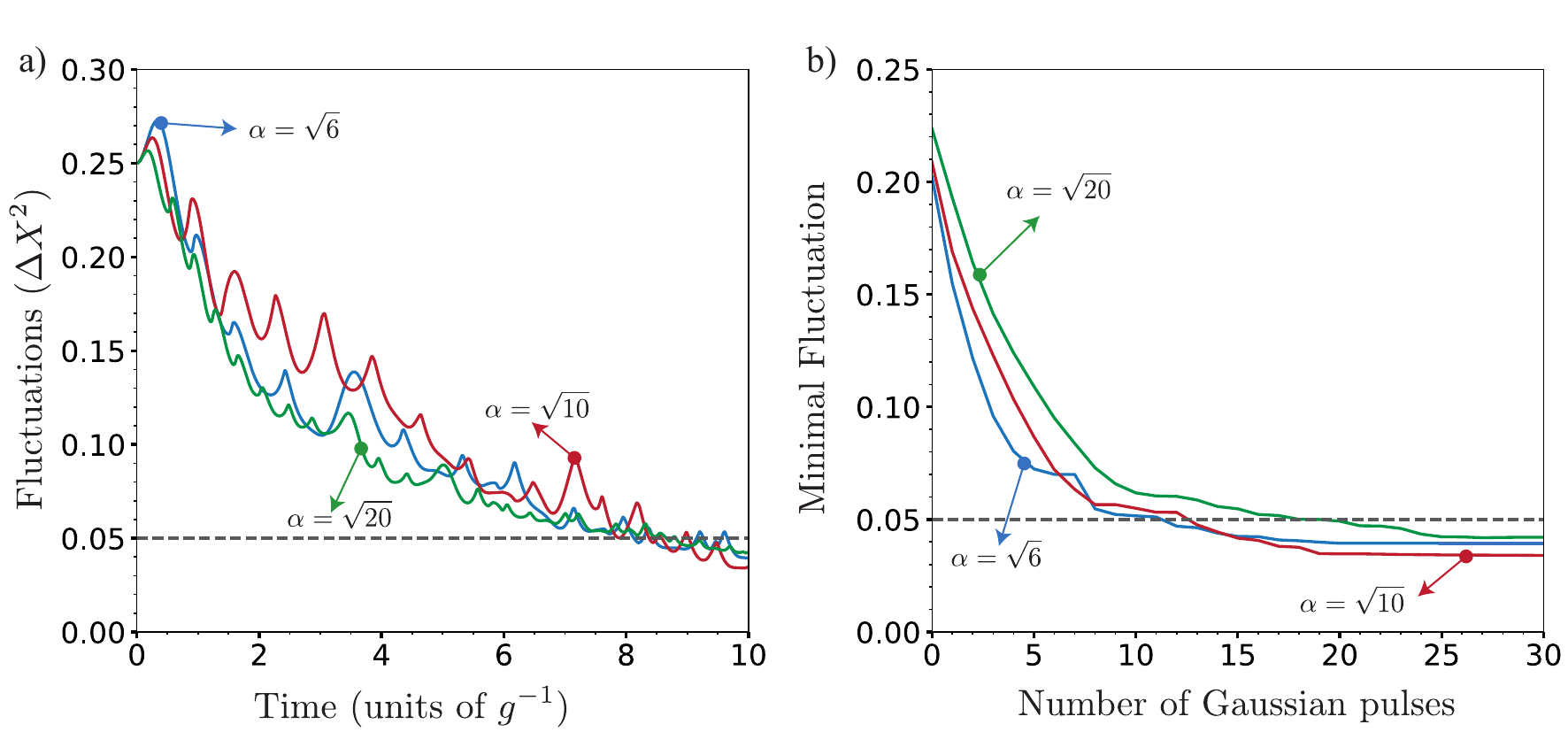}
	\caption{Fluctuation dynamics for coherent states with $\alpha=\sqrt{6}$, $\alpha=\sqrt{10}$, and $\alpha=\sqrt{20}$. We consider for the Gaussian pulses $g\sigma=0.025$. a) Time evolution of cuadrature fluctuation $\Delta X^2$ using 30 Gaussian pulses. b) Minimal value of fluctuation reached with different number of Gaussian pulses.} 
	\label{Fig05}
\end{figure}

\emph{Gradient-based iterative disorder strategy}.\textemdash In the GB-IDS, we follow the same procedure as in the IDS, but now, instead of defining a time grid and perform a brute force optimization, we use a gradient-base optimizer, specifically in this case, we use the \textit{trust-constr} algorithm from the \textit{SciPy} python library. In each iteration, we use the time we want to modify as a seed for the optimization algorithm. This issue allows the algorithm to speed up since, in each iteration, the modification of the times is not large. 

\emph{Gradient-free iterative disorder strategy}.\textemdash In the GF-IDS, we follow exactly the same procedure that in the GB-IDS and IDS, but now, instead of a gradient-base or brute force optimizer, we use a gradient-free algorithm specifically in this case, we use the \textit{COBYLA} from the \textit{SciPy} python library. We need to mention that we use the same method as in the GB-IDS to define the seed for the optimization algorithm.

Figure \ref{Fig02} shows the performance of each strategy for the initial state $\ket{e}\ket{\alpha}$, with $\alpha=\sqrt{6}$ and $g\sigma=0.05$. From this result, it is important to mention that the FSS finds only seven different pulses in the time windows $gt\in[0,10]$, and the other strategies can find 15 different pulses, allowing further reduction of quantum fluctuations. Also, in Fig.~\ref{Fig02}, we can see that IDS and IDS-GF find equivalent solutions and are better than the pulse train given by FSS and IDS-GB. It suggests that our cost functions, \,  i.e., the minimal fluctuation for a set of pulses, have several global minima, implying that a gradient-based optimizer cannot reach a global one. 

To give a quantitative value of the reduction, we can define the reduction percentage with respect to the shot noise limit ($\mathcal{R}\%$) as
\begin{equation}
\mathcal{R}\%=\frac{0.25-\Delta X^2}{0.25} \times 100\%
\label{reductionper}
\end{equation}
It is important to highlight that even if FSS gives the worst result, it is a very fast strategy and reaches a non-trivial reduction larger than $70\%$ related to the shot-noise limit. Using the other strategies, we obtain always $\mathcal{R}\%>80\%$. Specifically for IDS $\mathcal{R}\%>86.11\%$, for IDS-GF $\mathcal{R}\%>86.64\%$, and for IDS-GB $\mathcal{R}\%>83.31\%$. We note that IDS and IDS-GF have almost the same performance, but in our simulation, IDS is faster due to that the time grid necessary to get the result is not large. Actually, we use a coarse grain time grid (time step of $\delta t= 0.01g^{-1}$) and around the minima a fine grained time-grid (time step of $\delta t=0.001g^{-1}$) that allows us to reduce the number of iterations in brute-force optimization. In the following cases, we only consider the IDS due to the better performance with respect to the other strategies, that is similar results in smaller time.

We consider three different values for the width of the Gaussian pulses, that is, $g\sigma=0.1$, $g\sigma=0.05$  and $g\sigma=0.025$. In Fig.~\ref{Fig03}, we show the results considering Gaussian pulses with $g\sigma=0.1$ and initial state given by $\ket{\alpha}\ket{e}$ with $\alpha=\{\sqrt{6},\sqrt{10},\sqrt{20}\}$. Figure~\ref{Fig03} a) shows the time evolution of the fluctuations for the optimal control using 30 pulses using the IDS, where the maximal reduction is reached for $\alpha=\sqrt{10}$ with $\mathcal{R}\%=84.59\%$. Figure~\ref{Fig03} b) shows the minimal fluctuation obtained after the optimization process for different numbers of pulses. This figure shows that the maximal reduction is almost reached with $15$ pulses, for the case $\alpha=\sqrt{10}$, with $\mathcal{R}\%=84.54\%$. In contrast, for $\alpha=\sqrt{20}$, we obtain less fluctuation reduction, reaching only $\mathcal{R}\%=21.31\%$ using five Gaussian pulses, and remain unchanged when we add more pulses. It suggests that with this protocol, few pulses are needed to reach a maximal fluctuation reduction.

In the same line Fig.~\ref{Fig04} and Fig.~\ref{Fig05} show the results considering gaussian pulses with $g\sigma=0.05$ and  $g\sigma=0.025$ respectively for the same initial states  $\ket{\alpha}\ket{e}$ with $\alpha=\{\sqrt{6},\sqrt{10},\sqrt{20}\}$. Panel a) for both figures shows the time evolution of the fluctuations for the optimal control using 30 pulses using the IDS. With  $g\sigma=0.05$  the maximal reduction is reached for $\alpha=\sqrt{20}$ (see Fig.~\ref{Fig04}) with $\mathcal{R}\%=88.73\%$, while for $g\sigma=0.025$ is reached for $\alpha=\sqrt{10}$ (see Fig.~\ref{Fig05} with $\mathcal{R}\%=86.38\%$. Also, it is interesting that in both cases, that is $g\sigma=0.05$ and  $g\sigma=0.025$, all the cases under study have a fluctuation reduction larger than the $80\%$, being in a case close to $90\%$. Now, from panel b), again we can observe that fluctuation reduction does not change appreciably after a given number of pulses, for example, in the case of Fig.~\ref{Fig04} b), the curve for $\alpha=\sqrt{20}$ reach $\mathcal{R}\%=88.48\%$ with 28 pulses, and in the case of Fig.~\ref{Fig05} b) the curve for $\alpha=\sqrt{10}$ reach $\mathcal{R}\%=86.11\%$ with 20 pulses.

\begin{figure}[t]
	\centering
	\includegraphics[width=1\linewidth]{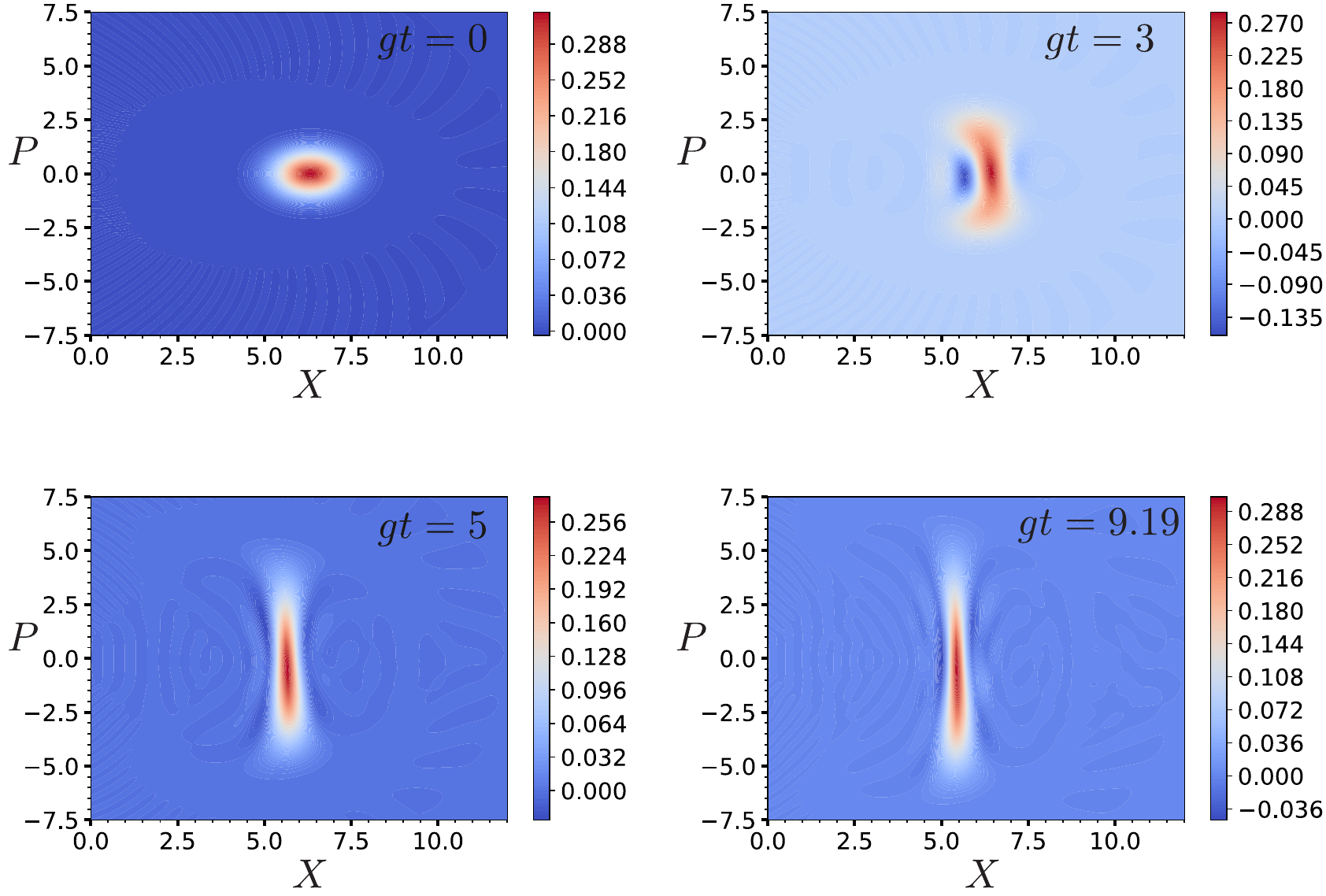}
	\caption{Wigner function for different times for the evolution given by 30 pulses using IDS, with $g\sigma=0.05$ and $\alpha=\sqrt{20}$.} 
	\label{Fig06}
\end{figure}

Finally, it is interesting to show how the Wigner function evolves under our protocol. Figure~\ref{Fig06} shows the Wigner function for the bosonic mode in different times for the case $g\sigma=0.05$ and $\alpha=\sqrt{20}$, that is, the maximal reduction obtained for our protocol. In $gt=0$ we have a coherent state, in $gt=3$ we already kick the system with seven pulses obtaining $\mathcal{R}\%=39.47\%$, later in $gt=5$ the system was driven by 12 pulses reaching $\mathcal{R}\%=70.18\%$. Finally, in $gt=9.19$, we obtain the maximal reduction after the 30 pulses $\mathcal{R}\%=88.73\%$.

\emph{Circuit QED implementation}.\textemdash Light-matter interaction in the microwave regime, known as circuit quantum electrodynamics (QED) \cite{Blais2021May}, is an ideal platform for realizing our noise reduction protocol. Here, it is possible to reach a large light-matter coupling strength that surpasses the TLS and cavity decay rates $g\gg\gamma,\kappa$. Our proposal considers the Jaynes-Cummings model, where a single TLS interacts with a single cavity mode. Considering the state-of-the-art superconducting circuits, a possible physical realization of our proposal may consider a fluxonium \cite{Manucharyan2009Oct,Nguyen2019Nov} that exhibits a large anharmonicity compared with the transmon \cite{Koch2007Oct,Schreier2008May}. Recent experiments have shown anharmonicities $(\omega_{20}-\omega_{10})/\omega_{10}$ as large as $17.85$ \cite{Ding2023Sep} and $1.771$ \cite{Bao2022Jun}.  Also, considering the lowest cavity mode of a coplanar waveguide or a lumped LC resonator, the Jaynes-Cummings model may describe the experimental situation at temperatures about $T=20$ mK. Antennas for fluxonium control are feasible within this technology, which leads to driving as $H_{\rm drive}=\hbar \Omega(t)q_F$, where $q_F$ is the fluxonium charge operator and $\Omega(t)$ is a Gaussian pulse (\ref{Gpulse}) which can be readily implemented in superconducting circuits  \cite{Ciani2022Nov,Krantz2019Jun}.  

\begin{figure}[b]
	\centering
	\includegraphics[width=0.8\linewidth]{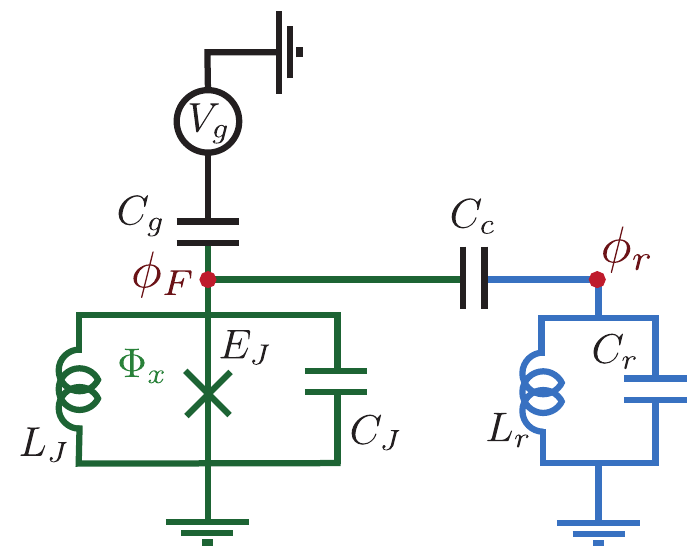}
	\caption{Equivalent circuit describing a fluxonium (green) capacitively coupled to an LC resonator (blue) and driving voltage source $V_g$ (black).} 
	\label{Fig07}
\end{figure}

Our model considers a fluxonium capacitively coupled to an LC resonator and driven by a voltage source $V_g$, as shown in the equivalent circuit of Fig.~\ref{Fig07}. The Hamiltonian that describes the circuit can be obtained using standard circuit quantization techniques \cite{Vool2017Jul}; see the Supplemental Material for a detailed derivation. In the limit $C_c\ll C_r,C_g$, the circuit Hamiltonian reads 
\begin{eqnarray}
H&=4E_CN_F^2+\frac{E_L}{2}\theta_F^2-E_J\cos(\theta_F+2\pi f)+4E_{C_r}N_r^2+\frac{E_{L_r}}{2}\theta_r^2\nonumber\\
&+2e\frac{C_g}{C_J+C_g}V_gN_F+2e\frac{C_g}{C_J+C_g}\frac{C_c}{C_r}V_gN_r+\frac{4e^2}{C_J+C_g}\frac{C_c}{C_r}N_FN_r.\nonumber\\
\end{eqnarray}

Here, $e$ is the electron charge, $E_C$, $E_L$, and $E_J$ are the fluxonium's charging, inductive, and Josephson energy, and we define the frustration parameter $f=\Phi_0/\Phi_x$, where $\Phi_0$ is the flux quantum and $\Phi_x$ is an external magnetic flux threading the fluxonium loop. $E_{C_r}$ and $E_{L_r}$ are the LC resonator's charging and inductive energy, respectively. Also, $N_\alpha$ and $\theta_\alpha$ ($\alpha=F,r$) are the number of Cooper pairs and phase-invariant gauge operators, respectively. All capacitances in the Hamiltonian are depicted in Fig.\ref{Fig07}.

It is worth noticing that the voltage source $V_g$ induces a remanent displacement on the LC resonator $\propto V_gN_r$. However, this driving is diminished by a factor $C_c/C_r$ compared to the fluxonium driving $\propto V_gN_F$ needed for our noise reduction protocol. Typical small capacitances in superconducting circuits may reach values of $C_c\simeq 1$fF \cite{Cicak2010Mar}, and resonator capacitances can readily reach values of $C_r\simeq 100$fF \cite{Forn-Diaz2016Jun}. Therefore, it is feasible within superconducting circuits technology to reach ratios $C_c/C_r\approx 10^{-2}$, and we can safely neglect undesired resonator displacement. 

Discussing the time scales involved in a circuit QED implementation is noteworthy. Considering a fluxonium such as in Ref.~\cite{Bao2022Jun} with transition frequency $\omega_{10}=2\pi\times 1.09$~GHz, coherence times $T_1\sim 40-120\mu s$ and $T_2\sim 10-35\mu s$, we estimate the light-matter coupling as $g=\omega_{10}/200\sim 2\pi\times 5$~MHz. Our proposed noise reduction protocol with Gaussian pulses occurs within the time scale $gt=10$, which implies a time of $t\sim 0.14 \mu$s to reach a reduction over $80\%$. This time is well below the coherence times in the current experimental situation. Therefore, we only consider numerical simulations of the unitary dynamics in this work.

\emph{Conclusion}.\textemdash We have proposed a protocol based on optimal control to reduce the quantum fluctuation in the bosonic mode of a light-matter system. The optimal control is done over the two-level system in a Jaynes-Cummings regime. Specifically, we found optimal Gaussian pulses train to reduce the quantum fluctuation more than the $88\%$. 

Also, we have experimentaly feasible considerations, for example, the use of Gaussian pulses over a two-level system is in the-state-of-the-art of any quantum computing platforms like traped ions, superconducting circuits, quantum dots, and more. Specifically, we perform the numerical simulations of our protocol considering a superconducting architecture based on a fluxonium coupled to a transmission line, obtaining that our proposal can be implementable with the current technology.

In conclusion, we have proposed an experimentally feasible protocol for noise reduction in a light-matter quantum system in a deterministic way, using optimal control techniques, providing a non-trivial level of squeezing useful for quantum technologies. Finally, this work paves the way for the use of optimal control to design protocols to enhance quantum features useful in quantum technologies.

\emph{Acknowledgments}.\textemdash We acknowledge financial support from Agencia Nacional de Investigaci\'on y Desarrollo (ANID): Financiamiento Basal para Centros Cient\'ificos y Tecnol\'ogicos de Excelencia (Grant No. AFB220001), Fondecyt grant 1231172, Subvenci\'on a la Instalaci\'on en la Academia grant SA77210018.

\bibliography{Mybib}

\newpage
\widetext
\section{Supplementary Material: Quantum control  and noise reduction of a quantized single mode and a two-level system} 
\date{\today}





\maketitle
\onecolumngrid
\section{Forward sequential strategy (FSS).}
\begin{algorithm}[H]
\caption{Forward sequential strategy (FSS)}
\begin{algorithmic}[1]
\State $H_{JC}$: Jaynes-Cummings Hamiltonian ($g=1$).
\State $\ket{\Phi(0)}=\ket{\alpha}\ket{e}$: Initial state.
\State Definition of the Gaussian pulse parameters: ($\sigma$, $\Omega_0$, $\omega_p$).
\State N: Total number of pulses.
\State $t_0=0$.
\For{$k_1$ from $1$ to N}
    \State Signal=0.
    \For{$k_2$ from $1$ to $k_1-1$}
    	\State Signal=Signal $+$ Gaussian pulse centered in $t_{k_2}$.
    \EndFor
    \State  T: equally spaced array from $t_{k_1-1}$ to $10$ with step $0.01$.
    \State  lT: number of elements of T.
     \For{$k_2$ from $1$ to lT}
    	\State $S_T$=Signal $+$ Gaussian pulse centered in T($k_2$) .
	\State Integrate the Schrodinger with $S_T$ as control function and initial state $\ket{\Phi(0)}$.
	\State Calculate $\Delta X^2(t)$ for $t\in [0,10]$.
	\State $X_{k_2}$: Minimal value of $\Delta X^2(t)$.
    \EndFor
    \State $k_2^*$: Index of the minimal value of all $X_{k_2}$.
    \State $t_{k_1}=T(k_2^*)$.
\EndFor
    \end{algorithmic}
\end{algorithm}

\newpage
\section{Iterative disorder strategy (IDS).}
\begin{algorithm}[H]
\caption{Iterative disorder strategy (IDS)}\label{AlgorithmIDS}
\begin{algorithmic}[1]
\State $H_{JC}$: Jaynes-Cummings Hamiltonian ($g=1$).
\State $\ket{\Phi(0)}=\ket{\alpha}\ket{e}$: Initial state.
\State Definition of the Gaussian pulse parameters: ($\sigma$, $\Omega_0$, $\omega_p$).
\State N: Total number of pulses.
\State $t_0=0$.
\For{$k_1$ from $1$ to N}
\State $\delta=10$.
\State $X_{aux}=10.$
\State $t_{k_1}=\textrm{rand}(0,10)$.

\While{$\delta>10^{-5}$}
    \State Signal=0.
    \For{$k_2$ from $1$ to $k_1$}
    \For{$k_3$ from $1$ to $k_1$; $k_3\ne k_2$}
    	\State Signal=Signal $+$ Gaussian pulse centered in $t_{k_3}$.
    \EndFor
    \State  T: equally spaced array from $0$ to $10$ with step $0.01$.
    \State  lT: number of elements of T.
     \For{$k_3$ from $1$ to lT}
    	\State $S_T$=Signal $+$ Gaussian pulse centered in T($k_3$) .
	\State Integrate the Schrodinger with $S_T$ as control function and initial state $\ket{\Phi(0)}$.
	\State Calculate $\Delta X^2(t)$ for $t\in [0,10]$.
	\State $X_{k_3}$: Minimal value of $\Delta X^2(t)$.
    \EndFor
    \State $k_3^*$: Index of the minimal value of all $X_{k_3}$.
   \State \textbf{Refinement process:}
    \State $t1=T(k_3^*-1)$ .
    \State $t2=T(k_3^*+1)$ .
    T: equally spaced array from $t1$ to $t2$ with step $0.001$.
     \State  lT: number of elements of T.
     \For{$k_3$ from $1$ to lT}
    	\State $S_T$=Signal $+$ Gaussian pulse centered in T($k_3$) .
	\State Integrate the Schrodinger with $S_T$ as control function and initial state $\ket{\Phi(0)}$.
	\State Calculate $\Delta X^2(t)$ for $t\in [0,10]$.
	\State $X_{k_3}$: Minimal value of $\Delta X^2(t)$.
    \EndFor
     \State $k_3^{**}$: Index of the minimal value of all $X_{k_3}$.
    \State $t_{k_2}=T(k_3^{**})$.
    
\EndFor
\State $\delta=|X_{aux}-X_{k_3^{**}}|$.
\State $X_{aux}=C$.
\EndWhile
\EndFor
\end{algorithmic}
\end{algorithm}

\newpage
\section{Gradient-based iterative disorder strategy (GB-IDS).}
\begin{algorithm}[H]
\caption{Gradient based iterative disorder strategy (GB-IDS)}\label{AlgorithmGBIDS}
\begin{algorithmic}[1]
\State $H_{JC}$: Jaynes-Cummings Hamiltonian ($g=1$).
\State $\ket{\Phi(0)}=\ket{\alpha}\ket{e}$: Initial state.
\State Definition of the Gaussian pulse parameters: ($\sigma$, $\Omega_0$, $\omega_p$).
\State N: Total number of pulses.
\State $t_0=0$.
\For{$k_1$ from $1$ to N}
\State $\delta=10$.
\State $X_{aux}=10.$
\State $t_{k_1}=\textrm{rand}(0,10)$.

\While{$\delta>10^{-5}$}
    \State Signal=0.
    \For{$k_2$ from $1$ to $k_1$}
    \For{$k_3$ from $1$ to $k_1$; $k_3\ne k_2$}
    	\State Signal=Signal $+$ Gaussian pulse centered in $t_{k_3}$.
    \EndFor
    \State \textbf{Define} Cost($\tau$) as:
    \State \indent $S_{\tau}$=Signal $+$ Gaussian pulse centered in $\tau$ .
    \State \indent Integrate the Schrodinger with $S_{\tau}$ as control function and initial state $\ket{\Phi(0)}$.
    \State \indent Calculate $\Delta X^2(t)$ for $t\in [0,10]$.
    \State \textbf{Return:}  minimal value of $\Delta X^2(t)$
\State \textbf{Minimize} CostFunction($\tau$) for $\tau\in[0,10]$ using a gradient based algorithm as \textit{trust-constr}.
    \State $t_{k_2}=\tau^*$ where Cost($\tau^*$) is the result of the minimization process.
    \State$C=$Cost($\tau^*$)
\EndFor
\State $\delta=|X_{aux}-C|$.
\State $X_{aux}=C$.
\EndWhile

\EndFor
\end{algorithmic}
\end{algorithm}

\newpage
\section{Gradient free iterative disorder strategy (GF-IDS).}
\begin{algorithm}[H]
\caption{Gradient based iterative disorder strategy (GF-IDS)}\label{AlgorithmGFIDS}
\begin{algorithmic}[1]
\State $H_{JC}$: Jaynes-Cummings Hamiltonian ($g=1$).
\State $\ket{\Phi(0)}=\ket{\alpha}\ket{e}$: Initial state.
\State Definition of the Gaussian pulse parameters: ($\sigma$, $\Omega_0$, $\omega_p$).
\State N: Total number of pulses.
\State $t_0=0$.
\For{$k_1$ from $1$ to N}
\State $\delta=10$.
\State $X_{aux}=10.$
\State $t_{k_1}=\textrm{rand}(0,10)$.

\While{$\delta>10^{-5}$}
    \State Signal=0.
    \For{$k_2$ from $1$ to $k_1$}
    \For{$k_3$ from $1$ to $k_1$; $k_3\ne k_2$}
    	\State Signal=Signal $+$ Gaussian pulse centered in $t_{k_3}$.
    \EndFor
    \State \textbf{Define} Cost($\tau$) as:
    \State \indent $S_{\tau}$=Signal $+$ Gaussian pulse centered in $\tau$ .
    \State \indent Integrate the Schrodinger with $S_{\tau}$ as control function and initial state $\ket{\Phi(0)}$.
    \State \indent Calculate $\Delta X^2(t)$ for $t\in [0,10]$.
    \State \textbf{Return:}  minimal value of $\Delta X^2(t)$
\State \textbf{Minimize} CostFunction($\tau$) for $\tau\in[0,10]$ using a gradient free algorithm as \textit{COBYLA}.
    \State $t_{k_2}=\tau^*$ where Cost($\tau^*$) is the result of the minimization process.
    \State$C=$Cost($\tau^*$)
\EndFor
\State $\delta=|X_{aux}-C|$.
\State $X_{aux}=C$.
\EndWhile

\EndFor
\end{algorithmic}
\end{algorithm}

\section{Hamiltonian derivation circuit QED implementation}
\begin{figure}[b]
	\centering
	\includegraphics[width=0.4\linewidth]{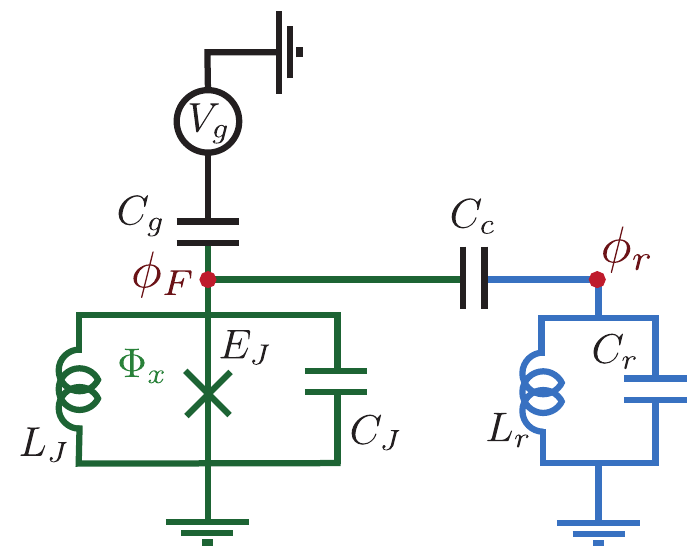}
	\caption{Equivalent circuit describing a fluxonium (green) capacitively coupled to an LC resonator (blue) and driving voltage source $V_g$ (black).} 
	\label{fig01}
\end{figure}

Our model considers a fluxonium capacitively coupled to an LC resonator, as shown in the equivalent circuit of Fig.~\ref{fig01}. The fluxonium consists of a capacitance $C_J$, a Josephson junction of energy $E_J$, and an inductor $L_J$, all in parallel. Also, $L_r$ and $C_r$ are the resonator's inductance and capacitance, respectively. An external voltage source $V_g$ is coupled to the fluxonium-resonator system via a capacitance $C_g$. The resonator and fluxonium are capacitively coupled via $C_c$. 

Defining actives nodes $\phi_F$ and $\phi_r$, the circuit Lagrangian may be written as follows
\begin{equation}
L=\frac{1}{2}\dot{\vec{\phi}}^T\mathcal{C}\dot{\vec{\phi}}-\dot{\vec{\phi}}^T\mathcal{C_G}\vec{V}_G-U(\phi_F,\phi_r), 
\end{equation}
where $\dot{\vec{\phi}}^T=(\dot{\phi}_F,\dot{\phi}_r)$, $\vec{V}_G^T=(V_g,0)$, and the potential energy $U(\phi_F,\phi_r)=\phi^2_F/(2L_J)-E_J\cos((\phi_F+\Phi_x)/\varphi_0)+\phi^2_r/(2L_r)$. Here, $\Phi_x$ is an external flux threading the fluxonium loop via a dedicated flux line. The matrix capacitances read
\begin{equation}
\mathcal{C}=\begin{pmatrix}
C_{\Sigma_F} & -C_c \\
-C_c & C_{\Sigma_r} 
\end{pmatrix},~ \mathcal{C_G}=\begin{pmatrix}
C_{g} & 0 \\
0 & 0 
\end{pmatrix},
\end{equation}
where the effective capacitances read $C_{\Sigma_F}=C_J+C_g+C_c$ and $C_{\Sigma_r}=C_r+C_c$. 

The Hamiltonian is obtained using the Legendre transformation $H=\vec{q}^T\dot{\vec{\phi}}-L(\vec{\phi},\dot{\vec{\phi}})$, where the conjugate momenta are contained in the vector $\vec{q}^T=(q_F,q_r)$. It can be shown that the circuit Hamiltonian reads
\begin{eqnarray}
H&=&\frac{1}{2}(\vec{q}+\mathcal{C_G}\vec{V}_G)^T\mathcal{C}^{-1}(\vec{q}+\mathcal{C_G}\vec{V}_G)+U(\phi_F,\phi_r)\nonumber\\
&=&\frac{C_{\Sigma_r}}{2|\mathcal{C}|}q_F^2+\frac{C_{\Sigma_r}C_g}{|\mathcal{C}|}V_g q_F+\frac{C_cC_g}{|\mathcal{C}|}V_g q_r+\frac{C_c}{|\mathcal{C}|}q_F q_r+\frac{C_{\Sigma_F}}{2|\mathcal{C}|}q_r^2 + U(\phi_F,\phi_r),
\end{eqnarray}
where $|\mathcal{C}|=(C_J+C_g)(C_c+C_r)+C_cC_r$ is the determinant of $\mathcal{C}$. In the limit $C_c\ll C_r,C_g$, $|\mathcal{C}|\approx(C_J+C_g)C_r$, $C_{\Sigma_F}\approx C_J+C_g$ and $C_{\Sigma_r}\approx C_r$. In this case, the circuit Hamiltonian reads
\begin{equation}
H=\frac{q_F^2}{2(C_J+C_g)}+\frac{C_g}{C_J+C_g}V_gq_F+\frac{C_g}{C_J+C_g}\bigg(\frac{C_c}{C_r}\bigg)V_gq_r+\frac{1}{C_J+C_g}\bigg(\frac{C_c}{C_r}\bigg)q_Fq_r+\frac{q_r^2}{2C_r} + U(\phi_F,\phi_r).
\end{equation}

Now, conjugate momenta $q_\alpha$ ($\alpha=F,r$) are related to the number of Cooper pairs $N_\alpha$ via the relation $q_\alpha=-2eN_\alpha$, where $e$ is the electron charge. In this way, we can rewrite the circuit Hamiltonian in terms of the number $N_\alpha$ and gauge-invariant phase $\theta_\alpha$ operators as follows
\begin{equation}
H = 4E_CN^2_F-\bigg(\frac{C_g}{C_J+C_g}\bigg)2eV_gN_F-\frac{C_g}{C_J+C_g}\bigg(\frac{C_c}{C_r}\bigg)2eV_gN_r + \frac{4e^2}{C_J+C_g}\bigg(\frac{C_c}{C_r}\bigg)N_FN_r + 4E_{C_r}N_r^2 +\frac{E_L}{2}\theta_F^2+\frac{E_{L_r}}{2}\theta_r^2-E_J\cos(\theta_F+2\pi f).
\label{Eq05}
\end{equation}

Here, $E_C=e^2/2(C_J+C_g)$, $E_L=\varphi_0^2/L_J$, and $E_J$ are the fluxonium's charging, inductive, and Josephson energy, while $E_{C_r}=e^2/2C_r$ and  $E_{L_r}=\varphi_0^2/L_r$ are the resonator's charge and inductive energy, and we define the frustration parameter $f=\Phi_0/\Phi_x$, where $\Phi_0$ is the flux quantum and $\Phi_x$ is an external magnetic flux threading the fluxonium loop. $E_{C_r}$ and $E_{L_r}$ are the LC resonator's charging and inductive energy, respectively. All capacitances in the Hamiltonian are depicted in Fig.\ref{fig01}. As is mentioned in the main text, typically $C_c\sim 1 \textrm{fF}$ and $C_r\sim 100 \textrm{fF}$, the third term in Eq.~(\ref{Eq05}) is neglectable respect to the second one. Then, the Hamiltonian can be approximated to 
\begin{equation}
H = \underbrace{4E_CN^2_F-\bigg(\frac{C_g}{C_J+C_g}\bigg)2eV_gN_F+\frac{E_L}{2}\theta_F^2-E_J\cos(\theta_F+2\pi f)}_{\textrm{Qubit}} + \overbrace{\frac{4e^2}{C_J+C_g}\bigg(\frac{C_c}{C_r}\bigg)N_FN_r}^{\textrm{Interaction}}+\underbrace{ 4E_{C_r}N_r^2 +\frac{E_{L_r}}{2}\theta_r^2}_{\textrm{Resonator}}.
\label{Eq06}
\end{equation}
The first four terms of the Hamiltonian are related to the fluxonium, the next one is related to the interaccion with the resonator, and the last two define the harmonic mode. We can write it in second quantization where 
\begin{eqnarray}
    N_F=\frac{i}{2}\left(\frac{E_L}{2E_C}\right)^{1/4}(b^{\dagger}-b), \quad N_r=\frac{i}{2}\left(\frac{E_{L_r}}{2E_{C_r}}\right)^{1/4}(a^{\dagger}-a)\nonumber\\
    \theta_F=\left(\frac{2E_C}{E_L}\right)^{1/4}(b^{\dagger}+b),\quad \theta_r=\left(\frac{2E_{C_r}}{E_{L_r}}\right)^{1/4}(a^{\dagger}+a)
\end{eqnarray}
As the fluxonium shows large anharmonicity (see ref.~[57] of the main text), we can diagonalize and truncate it to two levels, obtaining the quantum Rabi model Hamiltonian that describes a light-matter system
\begin{equation}
H = \hbar\frac{\omega_0}{2} \sigma_z + \hbar\omega a^{\dagger}a + i\hbar g \sigma_y(a^{\dagger}-a)
\label{Eq06}
\end{equation}
where $\omega_0\approx \sqrt{\frac{8E_L E_C}{\hbar^2}}$, $\omega=\sqrt{\frac{8E_{L_{r}}E_{C_{r}}}{\hbar^2}}$ and $g=\frac{C_c}{2C_r(C_J+C_g)}\sqrt{\frac{1}{Z_r Z}}$,  with $Z_r=\sqrt{\frac{L_r}{C_r}}$ and $Z=\sqrt{\frac{L_J}{C_J+C_g}}$ the impedances of the resonator and the fluxonium respectively.
\newpage
\section{Time Pulses}

\begin{table}[h]
\begin{tabular}{|c|c|c|c|c|}
  \hline
  & FSS & IDS & GB-IDS & GF-IDS \\
  \hline
  $t_1$ & 0.99 & 0.025 & 1.57664117 & 0.03333281 \\
  $t_2$ & 1.60 & 1.395 & 1.60009375 & 1.36789883\\ 
  $t_3$ & 2.36 & 2.205 & 1.66905474 & 2.1789207 \\ 
  $t_4$ & 3.17 & 3.001 & 2.25157374 & 2.98055742 \\ 
  $t_5$ & 3.99 & 3.775 & 2.25649523 & 3.75631523 \\ 
  $t_6$ & 4.29 & 4.601 & 2.36593208 & 4.5855125 \\ 
  $t_7$ & 5.09 & 6.415 & 2.37496655 & 6.40902734 \\ 
  $t_8$ & 5.21 & 7.384 & 2.38427474 & 7.3859375 \\ 
  $t_9$ & 5.26 & 8.344 & 2.61587942 & 8.37025469 \\ 
  $t_{10}$ & - & 8.596 & 3.40150903 & 8.66145039 \\ 
  $t_{11}$ & - & 8.759 & 4.2124007 & 8.82645117 \\ 
  $t_{12}$ & - & 9.144 & 5.01842212 & 9.22317539 \\ 
  $t_{13}$ & - & 9.291 & 5.34008877 & 9.36379375 \\ 
  $t_{14}$ & - & 9.609 & 5.56069759 & 9.64438477 \\ 
  $t_{15}$ & - & 9.999 & 5.73601587 & 9.99468281 \\ 
  $t_{16}$ & - & - & 6.14444342 & - \\ 
  $t_{17}$ & - & - & 6.31082951 & - \\ 
  $t_{18}$ & - & - & 6.52073723 & - \\ 
  $t_{19}$ & - & - & 7.00977128 & - \\ 
  $t_{20}$ & - & - & 7.66928555 & - \\ 
    \hline
\end{tabular}
\caption{Pulse times for different strategies obtained for $g\sigma=0.05$ and $\alpha=\sqrt{6}$. Data relevant for Figure 2 of the main text.}
\end{table}

\begin{table}[h]

\begin{tabular}{|c|c|c|c||c|c|c||c|c|c|}
  \hline
  & \multicolumn{3}{|c|}{$g\sigma=0.1$} & \multicolumn{3}{|c|}{$g\sigma=0.05$} & \multicolumn{3}{|c|}{$g\sigma=0.025$} \\
  \hline
  & $\alpha=\sqrt{6}$ & $\alpha=\sqrt{10}$ & $\alpha=\sqrt{20}$ & $\alpha=\sqrt{6}$ & $\alpha=\sqrt{10}$ & $\alpha=\sqrt{20}$ & $\alpha=\sqrt{6}$ & $\alpha=\sqrt{10}$ & $\alpha=\sqrt{20}$ \\
  \hline
  $t_1$ & 1.163 & 0.886 & 0.517 & 0.025 & 0.713& 0.582& 0.928&0.862 &0.556 \\
  $t_2$ & 1.327 & 1.360 & 0.909 & 1.395 & 1.166& 0.857& 1.55&1.401 &0.897 \\
  $t_3$ & 1.955 & 1.547 & 1.287 & 2.205 & 1.683& 1.205& 2.417&2.245 & 1.243\\
  $t_4$ & 2.844 & 1.945 & 1.649 & 3.001 &2.257 & 1.467& 4.338& 3.049& 1.620 \\
  $t_5$ & 3.601 & 2.651 & 2.630 & 3.775 &2.857 &1.873 & 5.325&3.856 & 2.031 \\
  $t_6$ & 4.402 & 3.259 & 2.716 & 4.601 & 3.482& 2.313& 5.665& 4.636& 2.473 \\
  $t_7$ & 5.120 & 3.856 & 2.799 & 6.415 & 4.945& 2.667&5.901 & 5.440& 2.863 \\
  $t_8$ & 5.886 & 4.450 & 2.841 & 7.384 &5.606 & 3.075& 6.169&5.956 & 3.934 \\
  $t_9$ & 6.600 & 5.036 & 2.847 & 8.344 & 6.314 & 3.471&7.150 & 6.167& 4.416 \\
  $t_{10}$ & 7.448 & 5.608 & 2.861  & 8.596 &6.626 & 3.874& 7.456& 6.474&5.575 \\
  $t_{11}$ & 8.445 & 6.154 & 2.861  & 8.759 & 7.250&4.29 &7.681 & 7.173& 5.783 \\
  $t_{12}$ & 8.446 & 6.806 & 2.865 & 9.144 & 7.280& 4.706& 7.941&7.592 &5.964 \\
  $t_{13}$ & 8.449 & 7.208 & 2.877 & 9.291 &7.290 & 5.749&8.202 &7.595 & 6.048 \\
  $t_{14}$ & 8.744 & 7.818 & 2.889 & 9.609 & 7.314& 6.298& 8.370& 8.199& 6.422\\ 
  $t_{15}$ & 8.889 & 9.000 & - & 9.999 & 7.345& 6.299& 9.202& 8.478& 7.026 \\
  $t_{16}$ & 9.460 & 9.060 & - & - & 7.360& 6.971&9.369 &8.568 & 7.202 \\
  $t_{17}$ & 9.460 & 9.067 & - & - &7.362 & 6.975& 9.474&8.988 & 7.843 \\
  $t_{18}$ & 9.461 & 9.089 & - & - & 7.379& 6.977&9.498 &9.268 & 8.018 \\
  $t_{19}$ & 9.594 & 9.089 & - & - &7.39 & 7.267& 9.604& 9.478&8.165 \\
  $t_{20}$ & 9.999 & 9.108 & - & - &7.396 &7.387 & 9.946&9.999 & 8.316 \\
    $t_{21}$ & - & 9.112 & -  &- & -& 7.530& 9.975&- &8.526  \\
  $t_{22}$ & - & 9.115 & - & -& -& 7.530& 9.980& -& 8.698 \\
  $t_{23}$ & - & 9.119 & - & -&- & 7.650& 9.992& -& 8.784 \\
  $t_{24}$ & - & 9.148 & - &- & -& 7.858& 9.998& -& 9.198 \\
  $t_{25}$ & - & 9.150 & - &- & -&8.117 & 9.999&- &9.698  \\
  $t_{26}$ & - & 9.150 & - &- & -&8.270 & -& -& 9.878 \\
  $t_{27}$ & - & 9.152 & - & - &- & 8.441&-&- & 9.937 \\
  $t_{28}$ & - & 9.156 & - & - &- & 8.719&- & -& 9.999 \\
  $t_{29}$ & - & 9.156 & - & - & -& 9.006& -& -& - \\
  $t_{30}$ & - & 9.161 & - & - &- &9.178 &- & -&-  \\
    \hline
    $\mathcal{R}\%$ &83.396&84.590&21.313&86.420&81.956&88.730&84.243&86.379&83.155 \\
      \hline
\end{tabular}
\caption{Pulse times for different coherent initial states and values for  $g\sigma$. Data relevant for Figure 3, Figure 4, and Figure 5 of the main text.}
\end{table}
\end{document}